# DISCOVERY OF POLARIZED LINE EMISSION IN SN1006

Sparks, W.B.[1], Pringle, J.E.[1,2], Carswell, R.F.[2], Long, K.S.[1], Cracraft, M.[1]

[1] Space Telescope Science Institute, 3700 San Martin Drive, Baltimore, MD 21218, USA.

[2] Institute of Astronomy, University of Cambridge, Madingley Road, Cambridge, CB3 0HA, UK.

Short title: POLARIZED LINE EMISSION IN SN1006

Corresponding author: sparks@stsci.edu

ABSTRACT

Laming (1990) predicted that the narrow Balmer line core of the ~3000 km s$^{-1}$ shock in the SN 1006 remnant would be significantly polarized due to electron and proton impact polarization. Here, based on deep spectrally resolved polarimetry obtained with the European Southern Observatory (ESO)'s Very Large Telescope (VLT), we report the discovery of polarized line emission of polarization degree 1.3% with position angle orthogonal to the SNR filament. Correcting for an unpolarized broad line component, the implied narrow line polarization is $\approx$2.0%, close to the predictions of Laming (1990). The predicted polarization is primarily sensitive to shock velocity and post-shock temperature equilibration. By measuring polarization for the SN1006 remnant, we validate and enable a new diagnostic that has important applications in a wide variety of astrophysical situations, such as shocks, intense radiation fields, high energy particle streams and conductive interfaces.

1. INTRODUCTION

The galactic supernova remnant SN 1006 is one of a number that show, in optical spectra, filaments dominated by the Balmer lines in neutral hydrogen (Kirshner, Winkler & Chevalier 1987; Raymond et al 1983). The filaments are thought to be thin sheets seen in projection, and are interpreted (Chevalier & Raymond 1978; see the review by Heng 2010) as high velocity (v > 200 km s$^{-1}$), non-radiative, collisionless shocks in partially ionized plasma. The Balmer radiation comes about through neutral hydrogen atoms passing through the shock and then interacting with the high temperature, post-shock plasma.

The Balmer emission profile has two main components (Chevalier, Kirshner & Raymond 1980; Morlino et al., 2012). There is a narrow component, with width of order 10 km s$^{-1}$ corresponding to the temperature of the pre-shock gas. This arises due to a small, non-zero probability that neutral atoms entering the shock are excited and then radiate before being ionized by the high-temperature, post-shock plasma. There is also a broad component, with width of order 1000 km s$^{-1}$, corresponding to the temperature of the post-shock gas. This is caused by charge exchange between the slow-moving neutral atoms and the rapidly moving post-shock protons producing fast-moving neutral atoms that are excited and radiate. The ratio between the flux in the narrow and broad components of the Balmer lines (usually discussed in the context of Hα) depends on the structure of the shock, and in particular on the rapidity with which the electron and proton temperatures are equilibrated at or downstream of the shock (Morlino et al 2012). Analysis of the Balmer-dominated spectra from non-radiative SNRs (Ghavamian, Laming & Rakowski, 2007) indicate that the ratio β of electron to proton temperature at the shock is close to unity for slower velocity shocks ($v_s < 400$ km s$^{-1}$) and that it decreases, approaching the theoretical minimum value of β= $m_e/m_p$ = 0.0005, at shock speeds above around 2,000 km s$^{-1}$.

Laming (1990) proposed a further important diagnostic for the post-shock equilibration in terms of the polarization of the Balmer emission lines. This polarization arises because the neutral hydrogen atoms entering the shock see, in their rest frame, an oncoming "beam" of shocked protons and electrons, which cause the excitation. Emission lines resulting from excitation by particles with a non-isotropic velocity distribution are polarized, with the polarization direction correlated with the direction of anisotropy (cf. Sparks et al 2014). As the electron temperature increases towards the proton temperature, due to equilibration, the velocity distribution of the electrons becomes much larger than the shock velocity (which is roughly the speed at which the neutral hydrogen atoms are travelling through the post-shock gas). As this happens, the electron velocity distribution appears less and less anisotropic, and so excitation by the electrons leads to the subsequent Balmer radiation being less and less polarized. As a corollary, the polarization is predicted to arise predominantly in the narrow component of the Balmer lines.

In his review article, Heng (2010) noted that no successful search for polarization in the narrow H$\alpha$ lines from Balmer-dominated supernova shocks has ever been reported. In this paper, we remedy this situation and report the discovery of polarized line emission associated with the narrow core of the H$\alpha$ line in the SN 1006 remnant.

2. OBSERVATIONS

We obtained VLT UT1 spectropolarimetry of the SN 1006 remnant emission filaments using FORS2 in long slit spectropolarimetric mode. On entering the instrument, light encounters a polarization slit mask, a rotatable half wave plate, a Wollaston prism to split the polarization

o- and e-beams, and a grism to disperse the light. The polarization mask is comprised of a series of 22 arcsec long segments designed to ensure that the dual polarization beams do not overlap on the detector. Here, we use only the single 22 arcsec segment centered on the SNR as illustrated in Fig 1. We used the 300V+10 grism to obtain spectra from ≈450 nm to ≈900 nm and the GG435 order sorting filter, with a wide 2 arcsec slit to maximize the amount of light gathered, oriented orthogonal to the filament in position angle -35°, Fig. 1.

Each half wave plate setting allows derivation of a single Stokes parameter, but to reduce systematics, the same Stokes parameter is observed with the beams reversed using rotation of the half wave plate. Hence four wave plate rotations are used to yield a complete set of linear Stokes polarization spectra. Our observations used half wave plate rotation angles of 0, 22.5, 45, and 67.5 degrees. The spectral window includes the first three lines of the Balmer series, H$\alpha$ (656.3 nm), H$\beta$ (486.1 nm) and H$\gamma$ (434.1 nm), Fig. 2.

Observations were obtained in VLT service mode on the nights of 2013 April 12/13 (6 sequences), 2013 April 19/20 (8 sequences), and 2013 May 8 (4 sequences), for a total of 18 sequences, each sequence comprising a 305 s exposure for each of four retarder settings, yielding a total exposure time on target of 6.1 hrs. Both polarized and unpolarized standard star observations were provided by ESO, using the same observing procedure.

The data were debiassed, flat-fielded using a pixel sensitivity flat field, corrected for image shear following Sparks et al (2014), and co-added to result in a final set of (ordinary) o- and (extraordinary) e-beam pairs for each of the four retarder settings. A simple cosmic ray rejection algorithm was applied during the co-addition by comparing each frame, normalized by the median in the spatial direction, to a median of all 18 similar spectra. The cosmic rays were identified in a mask image, and omitted during the subsequent averaging of like-frames. To derive the polarization information, we used the flux ratio method (Miller et al. 1987).

The normalized Stokes parameters are given by $q = (R_q - 1)/(R_q + 1)$, where $R_q = \sqrt{\left(\frac{I_0^o}{I_0^e}\right)/\left(\frac{I_{45}^o}{I_{45}^e}\right)}$ and $u = (R_u - 1)/(R_u + 1)$, where $R_u = \sqrt{\left(\frac{I_{22.5}^o}{I_{22.5}^e}\right)/\left(\frac{I_{67.5}^o}{I_{67.5}^e}\right)}$. The Stokes I and normalized $q$ and $u$ frames were derived without sky subtraction, which resulted in a clean set of $q$ and $u$ images since sky lines are unpolarized, and a Stokes *I* image which includes the sky. The normalized $q$ and $u$ data were converted to polarized intensities Stokes *Q* and *U* by multiplying $q$ and $u$ by the *I* data. We then subtracted a sky estimate from the Stokes *I* image by averaging the spectra below the bright Hα rim seen in Fig. 2 (to the lower left in Fig. 1), and subtracting it from Stokes *I*. This worked well, as is evident from the lower panel of Fig. 2. To measure the line emission polarization we took the 2.5 arcsec wide region covering the bright rim (the entire filament structure is ≈10 arcsec across), and derived a spatially integrated spectrum for *I*, *Q* and *U*, Fig. 3. We integrated the section from 653.6 to 658.6 nm for Hα and 483.6 to 488.6 for Hβ, after subtracting "continuum" regions either side, 619.6–649.6 nm and 662.6–692.6 nm for Hα and 449.6–479.6 and 492.6–522.6 nm for Hβ. The continuum baseline subtraction served to remove any zero point offsets from the *I, Q, U* intensity spectra. The mean values for baseline subtracted *Q* and *U* were divided by the mean of the sky subtracted, baseline subtracted Stokes *I* to derive new, final normalized Stokes parameters $q_f$ and $u_f$, and hence polarization degree $p = \sqrt{q_f^2 + u_f^2}$, and position angle $\theta = \frac{1}{2}\tan^{-1}(U/Q) - \phi$ where $\phi$ includes the instrument rotation on the sky to slit position angle -35° and the retarder offset calibration provided by ESO[1]. The rms dispersion of *Q* and *U* about the baseline fits provided the uncertainty estimates used for Table 1.

---

[1] http://www.eso.org/sci/facilities/paranal/instruments/fors/inst/pola.html

## 3. RESULTS

Fig. 3 illustrates the data used to derive the polarimetric results presented in Table 1. For the narrow core of the dominant H$\alpha$ line, polarization is detected at a level $\approx$ 4.5 $\sigma$, $p\approx$0.0128 polarization degree (i.e. 1.3% polarization) and position angle at 143.7°, only 1.3° from the perpendicular to the filament. The H$\beta$ polarization is barely significant, at a $\approx$ 2 $\sigma$ level, though the values are consistent within the uncertainties with those for H$\alpha$, at position angle 13°+/-13° from the filament perpendicular. The H$\alpha$ results form the basis of our assertion that polarized line emission has been discovered in the SN1006 remnant. We estimated that the intensity of the narrow component $I_n$ to that of the broad component $I_b$, *within the spectral range of the narrow component only*, is $I_n/I_b\approx$1.9 (cf. Nikolic et al 2013). Hence if only the narrow component is contributing to the polarization, the implied corrected polarization degree of the narrow component is $p_n\approx$0.0197, i.e. $\approx$2.0% orthogonal to the filament direction.

## 4. DISCUSSION

Laming (1990) predicted emission line polarization due to the highly anisotropic impact of energetic protons and electrons. Specifically, Laming (1990) considers strong shocks in a pure hydrogen plasma. For the range of shock parameters he considers, he finds that the polarization vector should be normal to the plane of the shock front. His analysis gives predicted values of the polarization of the narrow H$\alpha$ component as seen in a plane perpendicular to the direction of motion of the shock front, for various values of the shock velocity and of the ratio $\beta = T_e/T_i$ of the electron to proton post-shock temperatures. He also gives two sets of values, depending on whether the Ly$\beta$ transition is optically thin (Case A) or optically thick (Case B). The difference here is that if Ly$\beta$ photons are absorbed and then re-emitted at Hα, then those Hα photons would have essentially zero net polarization.

Ghavamian et al (2002) model the optical spectra of SN1006. They observe essentially the same filament section as we do. From their analysis they conclude that the shock velocity is 2890 +/- 100 km s$^{-1}$, and they require a low degree of electron-proton equilibration at the shock front, with β< 0.07. They also conclude that the Lyβ optical depth is low τ(Lyβ) ~ 0.5. Morlino et al (2012) consider Balmer emission from collisionless shocks in a partially ionized medium. For a shock velocity of 3,000 km s$^{-1}$ they find that the ratio of flux from the narrow and broad components ($I_n/I_b$) varies non-monotonically with β and lies in the range 0.6 to 2, taking the uppermost value when β ~ 0.01. This is consistent with our finding that $I_n/I_b$ ~ 1.2.

For Case A (low Lyβ optical depth), and for shock velocity of 3,000 km s$^{-1}$, Laming's (1990) analysis predicts that the polarization of the narrow Hα component should have a polarization in the range ≳2.7 to 6.5 percent, taking the lowest value when β ~ 0.01. For $I_n/I_b$ ~ 2, this would lead to a polarization of the Hα line as a whole of approximately ≳1.8 per cent, or ≳2.7% in the narrow core. Our measurement of 1.3% in the core includes a contribution from the broad component which we estimate to be ≈35% of the total, yielding a corrected observed polarization of ≈2.0%. Practical issues such as the dilution of polarization through an admixture of geometric effects and the addition of theoretical complications (non-hydrogen plasma; shock precursors; magnetic fields; cosmic rays) have not been considered, yet the empirical result of ≈2.0% polarization is encouragingly close to Laming's (1990) theoretical value of ≳2.7%.

## 5. CONCLUSION

In this work we confirm the prediction (Laming 1990) that the Balmer lines in non-radiative shocks in SNR are polarized, and that the direction of polarization is normal to the plane of the shock front. The formal analysis gives the angle between the polarization vector and the shock front normal as 1.3 +/- 6.3 degrees for H$\alpha$ and 13 +/- 13 degrees for H$\beta$, orthogonal to the shock front within the uncertainties. We find percentage polarizations of 1.3 +/- 0.3 percent for H$\alpha$ and 1.8 +/- 0.9 percent for H$\beta$. Correcting for an unpolarized broad component we find a polarization level of 2.0 +/- 0.4 percent for H$\alpha$. The fact that these values are slightly lower than those predicted in Laming's (1990) analysis is not unreasonable, given the limitations of that analysis already noted by Laming, and practical considerations mentioned above. However the basic, fundamental agreement underlines the importance of such polarization observations for our understanding of the physics of SNR shocks, and related astrophysical situations involving highly anisotropic excitation mechanisms, such as conduction fronts, high energy particle streams and intense radiation fields.

ACKNOWLEDGMENTS

We thank J.M. Laming for helpful remarks. Based on observations collected at the European Organisation for Astronomical Research in the Southern Hemisphere under ESO programme 091.D-0176(A) and in part on observations with the NASA/ESA *Hubble Space Telescope* which is operated by the Association of Universities for Research in Astronomy, Inc., under NASA contract NAS5-26555. Support for this work was provided by NASA through grants GO-12962 from the Space Telescope Science Institute, which is operated by the Association of Universities for Research in Astronomy, Inc., under NASA contract NAS5-26555. JEP is grateful for support from the Distinguished Visitor Program at STScI.

TABLES

Table 1. Polarimetry results for cores of Hα and Hβ lines in SN1006 remnant.

| SN1006 | Stokes I | $Q$ | $U$ | $p, p_n$* | $\theta$ |
|---|---|---|---|---|---|
| Hα | 52201.2 | 663.4 +/- 147.7 | -64.9 +/- 139 | 0.0128 +/- 0.0028<br>0.0197 +/- 0.0043 | 143.7 +/- 6.3 |
| Hβ | 12522.1 | 217.6 +/- 107 | -79.8 +/- 107 | 0.0185 +/- 0.0086 | 131.4 +/- 13.2 |

*The polarization of the total narrow-line core is $p$, and the estimated polarization of the narrow component only, $p_n$, assuming the broad component is unpolarized.

FIGURES

Figure 1. Location of 11 arcsec polarization slit segment on *Hubble Space Telescope* (HST) image of SN1006 remnant. North is up, East to the left; the SNR filament is in position angle ≈55°.

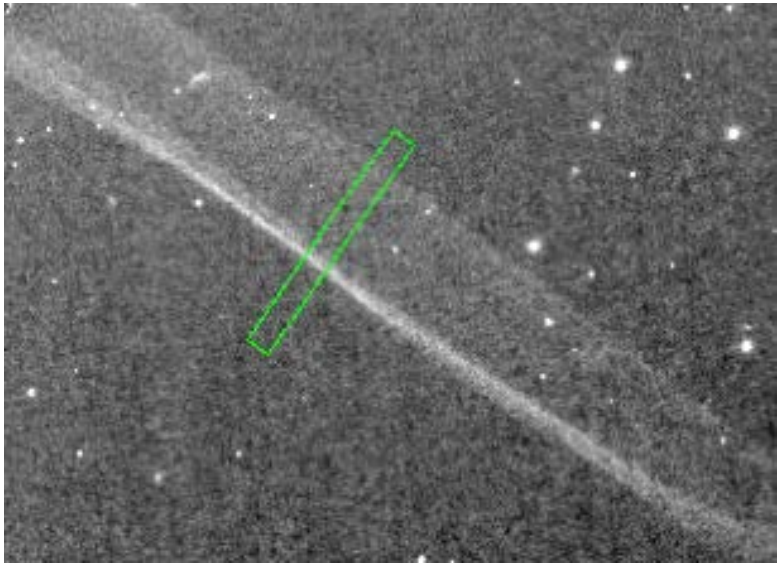

Figure 2. Upper, co-added spectra showing total intensity after basic reductions, and lower, same after subtraction of sky background showing strong Balmer dominated shock spectrum. Lines visible are, right to left, Hα, Hβ and Hγ.

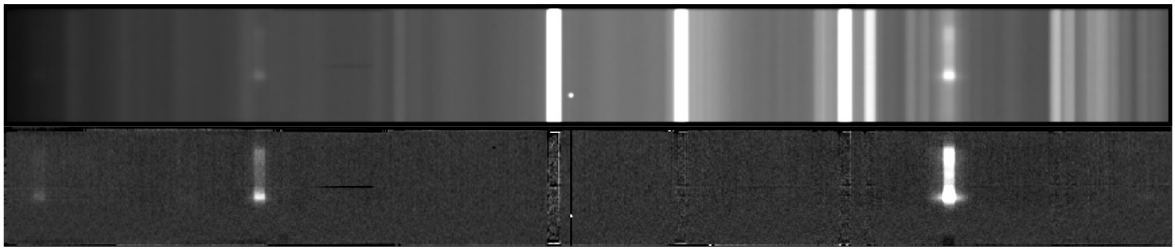

Figure 3. Un-normalized spectra for Stokes *I, Q, U*, each the total across a 2.5 arcsec spatial window centered on the bright filament rim, and averaged over 18 individual sequences. Hence the total counts are ≈18× the *y* axis values. The Q and U plots have been lightly

smoothed with a Gaussian σ≈0.37 nm.

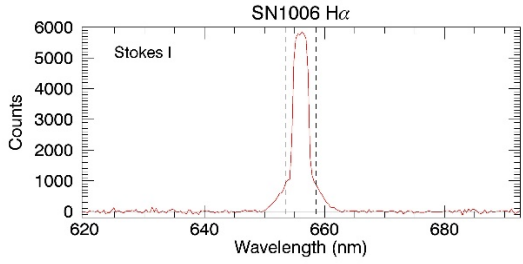
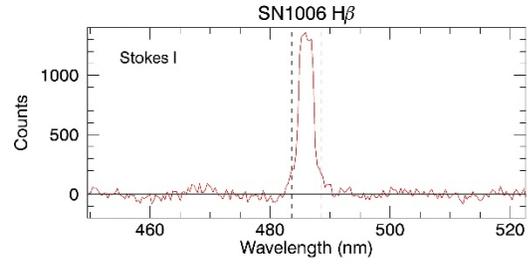
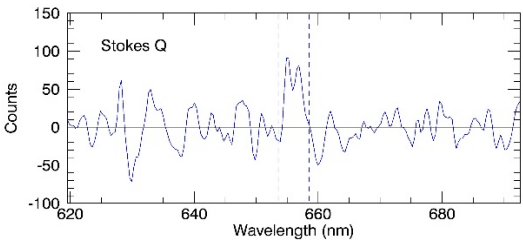
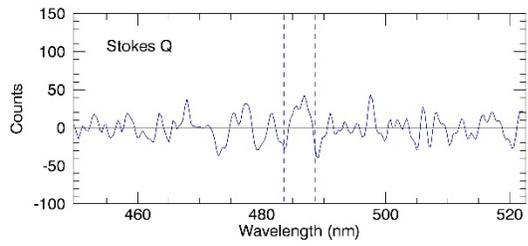
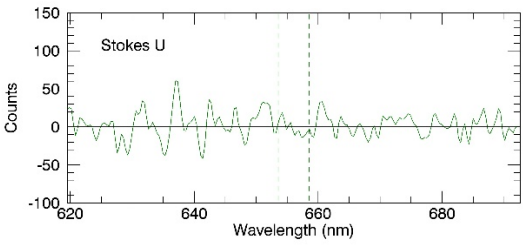
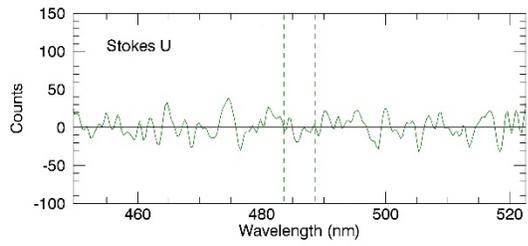